\documentclass[12pt,twoside]{article}
\usepackage[T1]{fontenc}
\usepackage[backend=bibtex,style=numeric-comp,autocite=plain,sorting=none,%
giveninits=true]{biblatex}
\usepackage{amsmath,amsfonts,amssymb,amsthm,mathabx}
\usepackage{times}
\usepackage{url}
\usepackage{tensor}
\usepackage{color}
\usepackage{caption}
\usepackage{enumitem}
\usepackage{graphicx}
\usepackage{array}
\usepackage{amsbsy}
\usepackage{appendix}
\usepackage{physics}
\usepackage{cancel}
\usepackage{yfonts}
\usepackage{chngcntr}
\usepackage{xcolor}
\usepackage{soul}
\usepackage{cancel}
\usepackage{setspace} 

\usepackage{hyperref}
\advance\textheight\topskip%
\numberwithin{equation}{section} \makeatletter
\@addtoreset{equation}{section}

\begin{document}

\title{Notes on non-equilibrium thermodynamics of Robinson-Trautman
  spacetimes}

\author{Glenn Barnich}


\def\mytitle{Notes on non-equilibrium thermodynamics of Robinson-Trautman
  spacetimes}

\pagestyle{myheadings} \markboth{\textsc{\small G.~Barnich}}
{\textsc{\small RT non-equilibrium thermodynamics}}

\addtolength{\headsep}{4pt}

\begin{centering}

  \vspace{1cm}

  \textbf{\Large{\mytitle}}

    \vspace{1.5cm}

    {\large Glenn Barnich}

\vspace{.5cm}

\begin{minipage}{.9\textwidth}\small \it \begin{center}
    Theoretical and Mathematical Physics \\
    Universit\'e libre de Bruxelles and International Solvay Institutes\\
    Campus Plaine C.P. 231,
    B-1050 Brussels, Belgium \\
    E-mail: \href{mailto:Glenn.Barnich@ulb.be}{Glenn.Barnich@ulb.be}
    \end{center}
  \end{minipage}

\end{centering}

\vspace{1cm}
  
\begin{center}
  \begin{minipage}{.9\textwidth} \textsc{Abstract}. Robinson-Trautman spacetimes are
    determined by a parabolic heat-type evolution equation of fourth order for a
    scalar field on future null infinity. This equation may be re-written as an
    energy balance for an asymptotically time-like observer. In turn, it becomes
    a first law in the context of non-equilibrium thermodynamics if the
    non-linear memory is interpreted as heat. This leads to a local version of
    temperature and a manifestly increasing entropy density from the viewpoint
    of the environment.
 \end{minipage}
\end{center}

\vfill
\thispagestyle{empty}

\newpage

\begin{spacing}{0.85}
    \tableofcontents
\end{spacing}

\newpage

\section{Introduction}\label{sec:introduction}

The RT evolution equation~\cite{Robinson:1960zzb}\cite{Robinson:1962zz}, which is equivalent to a two dimensional
Calabi flow~\cite{Tod1989}, arises in the context of a class of solutions to Einstein's
vacuum equations describing an isolated system settling down to boosted
Schwarzschild black holes by the emission of gravitational radiation. This
interpretation provides a definition of energy density and a Lyapunov function,
the Bondi supermomentum that can only decrease on account of the news.
Furthermore, it identifies the time-independent equilibrium sector with boosted
Schwarzschild black holes, whose equilibrium thermodynamics can be worked out by
standard methods.

Independently of further considerations from general relativity, our aim here is
to provide an interpretation of this evolution equation as a non-equilibrium
thermodynamic system. 

A unified approach to non-equilibrium thermodynamics does not seem to exist, see
e.g.~the collection of contributions in~\cite{Greven2014}
and~\cite{gaspard2025thermodynamicsirreversibleprocessesdissipative}. We will
follow the overview~\cite{Mueller2003},
\begin{itemize}
 \item the 1st law states the balance of energy
  \begin{equation}
    \label{eq:59}
    \frac{d (U+K)}{dt}= \dot Q+\dot A,
  \end{equation}
        where the rate of change of the sum of the internal and the kinetic
        energy is equal to the sum of the heating and the working, with the
        understanding that $\dot Q$ and $\dot A$ are not time derivatives but
        rates of exchange of heat and work, and that $Q$ is the time integral of
        $\dot Q$ over the duration of a heat exchange, while in simple
        thermodynamic applications, 
        \begin{equation}
          \label{eq:65}
          \dot A=-p\frac{dV}{dt}.
        \end{equation}

  \item the 2nd law states that the rate of change of entropy $S$ satisfies the
  inequality
  \begin{equation}
    \label{eq:61}
    \frac{d S}{dt}\geq \frac{\dot Q}{T}, 
  \end{equation}
  for irreversible processes, with equality holding for reversible ones.

  \item for reversible processes, elimination of $\dot Q$ yields the Gibbs
        relation
  \begin{equation}
    \label{eq:64}
    \frac{dS}{dt}=\frac{1}{T}\big(\frac{d (U+K)}{dt}-\dot A\big). 
  \end{equation}
        For simple thermodynamic applications where kinetic energy can be
        neglected, this reduces to
  \begin{equation}
    \label{eq:66}
    \frac{dS}{dt}=\frac{1}{T}\big(\frac{d U}{dt}+p\frac{dV}{dt}\big).
  \end{equation}
\end{itemize}
As done previously in the gravitational context, we will proceed by analogy and
identify suitable densities below.

As concerns Robinson-Trautman solutions and the associated equilibrium
thermodynamics, we follow the analysis and conventions of~\cite{Barnich:2026wpw,Barnich:2026dka}, to which
we refer for more details.

In the next section, we will start by reviewing the RT evolution equation and
its symmetries before discussing balance laws, including the generalized mass
aspect that leads to an improved Lyapunov functional. We conclude the section by
reviewing the equilibrium thermodynamics.

In the main Section~\ref{sec:non-equil-entr}, we follow the logic on non-equilibrium thermodynamics
as summarized in the introduction. Starting from the first law in terms of an
energy balance law, we identify heat in terms of the non-linear memory
associated to RT solutions. When using the second law for reversible processes,
this allows us to identify a local temperature together with an entropy density
from the viewpoint of the environment that does not decrease along the flow and
may be used to recast the RT equation as a local Gibbs relation.

Section~\ref{sec:phase-space-its} and Appendix~\ref{sec:dirac-brack-equil} are devoted to
complementary considerations on a phase space description of the system with a
Lorentz covariant separation of the degrees of freedom. At this stage it is not
entirely clear if such an understanding is indeed useful for the thermodynamic
problem at hand or if it may be used as a stepping stone towards a microscopic
description.

In the final Section~\ref{sec:discussion}, we compare the approach adopted here to earlier
results on non-equilibrium thermodynamics of RT spacetimes, either in context of
the fluid gravity or the AdS/CFT correspondence, and more generally to more
recent considerations on non-equilibrium thermodynamics of gravitational systems
associated to various types of horizons and null surfaces.

\section{RT equation: symmetries, balance laws, and equilibrium thermodynamics}\label{sec:rt-equat-symm}

\subsection{Robinson-Trautman evolution equation}\label{sec:robins-trautm-evol}

Let $V_\circ(u,x)$ be a positive function on the positive real line times the sphere
$R^*_+\times\mathbb S^2$, with coordinates $u,x^A=(\zeta,\widebar\zeta)$ and
$x^A=\zeta,\widebar\zeta$, where $\zeta=\cot\frac{\theta}{2}e^{\imath\phi}$ are stereographic
coordinates. The non-linear evolution equation that we are interested in is the
Robinson-Trautman (RT) equation,
\begin{equation}
  \label{eq:77}
  3\partial_u V_\circ=-V_\circ^4\eth^2_\circ\widebar\eth_\circ^2V_\circ+V_\circ^3|\eth_\circ^2 V_\circ|^2.
\end{equation}
In this equation, we use the ``eth'' operator on the sphere. If $\eta^s$ is a
field on the sphere of spin weight $s$, and
\begin{equation}
  \label{eq:101}
  P_\circ=\frac{|\zeta|^2+1}{\sqrt 2},
\end{equation}
the $\eth_\circ$ operator and its complex
conjugate are defined by
\begin{equation}
  \label{eq:5}
    \eth_\circ \eta^{s}=P^{1-s}_\circ\widebar\partial ( P^s_\circ \eta^s),\quad
    \widebar\eth_\circ\eta^s=P^{1+s}_\circ\partial (P^{-s}\eta^s),
\end{equation}
with $\partial=\partial_\zeta,\widebar\partial=\partial_{\widebar\zeta}$. In
particular, $V_\circ$ is of spin weight $0$ and
\begin{equation}
  \label{eq:115}
  \eth^2_\circ\widebar\eth_\circ^2 \eta^0=\frac 14 \Delta_\circ(\Delta_\circ +2)\eta^0,
\end{equation}
where $\Delta_\circ$ is the scalar Laplacian on the sphere. If one expands in terms of
spherical harmonics,
\begin{equation}
  \label{eq:122}
  V_\circ=q^{jm}(u) Y_{jm},\quad \widebar q^{jm}=(-)^m q^{j,-m}. 
\end{equation}

In summary, the Robinson-Trautman equation is a parabolic evolution equation,
akin to heat-type diffusion equations, for a real scalar field $V_\circ$ on
$\mathcal J^+$.

\subsection{Symmetries}\label{sec:symmetries}

Defining $u=\lambda^{-4}u^\prime$, $\lambda>0$, the scaling transformation
\begin{equation}
  \label{eq:138}
  V^\prime_{\circ}(u^\prime,x)=\lambda^{-1}V_{\circ}(\lambda^{-4}u^\prime,x), 
\end{equation}
maps solutions to solutions. Note that
\begin{equation}
  \label{eq:118}
  \partial_{u^\prime}V^\prime_{\circ}(u^\prime,x)=\lambda^{-5}(\partial_u V_\circ)(\lambda^{-4}u^\prime,x),
\end{equation}
which implies the covariance of equation~\eqref{eq:77} under the scaling
transformation.

A $\mathrm{SL}(2,\mathbb C)$ transformation $g^{-1}$ described by
$x_{g}(x^\prime)$,
\begin{equation}
  \label{eq:82}
x_{g}(x^\prime):\  \zeta=\frac{a\zeta'+b}{c\zeta'+d},\ \widebar \zeta=\frac{\widebar a\widebar \zeta'+\widebar b}{\widebar c\widebar \zeta'+\widebar d},
\end{equation}
acts on spin and boost weighted ``functions'' $\eta^{s,w}(x)$ of weights $(s,w)$ as
\begin{equation}
  \label{eq:347}
  \eta^{\prime s,w}_{;g^{-1}}(x^\prime)=e^{-w E^g_R-s\imath E^g_I}(x^\prime)\eta^{s,w}(x_{g}(x^\prime)),
\end{equation}
where
\begin{equation}
  \label{eq:153}
  e^{E^g_R}(x)=\frac{|\zeta|^2+1}{|a\zeta+b|^2+|c\zeta+d|^2},\quad e^{\imath E_I^g}(x)=\frac{\widebar c\widebar\zeta+\widebar d}{c\zeta+d}.
\end{equation}
When taking into account that $V_\circ$ is of spin and
boost weights $(0,1)$, 
\begin{equation}
  \label{eq:357}
  V^\prime_{\circ;g^{-1}}(u,x^\prime)
  =e^{-E^g_R}(x^\prime)V_{\circ}(u,x_g(x^\prime)).
\end{equation}
solutions are mapped to solutions.

There is a further discrete symmetry of solutions under the antipodal map.

\subsection{Balance laws for RT equation}
\label{sec:balance-laws-rt}

Two other equivalent forms of the RT evolution equation~\eqref{eq:77} are,
\begin{equation}
  \label{eq:104}
  \partial_u V_\circ^{-3}-\eth^2_\circ\widebar\eth^2_\circ V_\circ=-V_\circ^{-1}|\eth_\circ^2 V_\circ|^2,
\end{equation}
and 
\begin{equation}
  \label{eq:293}
  \frac 12 \partial_u V^{-2}_\circ+\frac 16 \eth_\circ\big(\eth_\circ V_\circ\widebar\eth_\circ^2V_\circ -V_\circ \eth_\circ \widebar\eth_\circ^2V_\circ
  \big)+\frac 16 \widebar\eth_\circ\big(\widebar\eth_\circ V_\circ\eth_\circ^2V_\circ  -V_\circ \widebar\eth_\circ \eth_\circ^2V_\circ  \big)=0.
\end{equation}

From gravitational considerations, it follows that the total energy and the energy density of the
system are the Bondi mass and mass aspect,
\begin{equation}
  \label{eq:99}
  E[V_\circ]=\int \frac{d^2\Omega}{4\pi}\ \varepsilon ,\quad \varepsilon= V_\circ^{-3}, 
\end{equation}
Re-writing~\eqref{eq:104}, the balance law for the energy density is
\begin{equation}
  \label{eq:100}
\partial_u \varepsilon + \eth_\circ \mathcal J_e+\widebar\eth_\circ\widebar{\mathcal J}_e=\sigma_e,
\end{equation}
in terms of the energy current
\begin{equation}
  \label{eq:105}
  \mathcal J_e=-\frac 12 \eth_\circ\widebar\eth^2_\circ V_\circ,
\end{equation}
and with energy-loss density
\begin{equation}
  \label{eq:107}
  \sigma_{e}=-V_\circ^{-1}|\eth_\circ^2 V_\circ|^2\leq 0.
\end{equation}

The local linear momentum density is 
\begin{equation}
  \label{eq:25}
   \pi_i=  V^{-3}_\circ(-n_i),
\end{equation}
with evolution equation
\begin{equation}
  \label{eq:26}
    \partial_u\pi_i-\eth_\circ^2\widebar\eth_\circ^2 V_\circ(-n_i)=- V_\circ^{-1}|\eth_\circ^2V_\circ|^2(-n_i). 
\end{equation}
This can also be written as a balance equation,
\begin{equation}
  \label{eq:131}
  \partial_u\pi_i +\eth_\circ\mathcal J_i+\widebar\eth_\circ \widebar{\mathcal J}_i=\sigma_i,
\end{equation}
where the linear momentum current and the linear momentum-loss density are
\begin{equation}
  \label{eq:6}
  \mathcal J_i=-\eth_\circ \widebar\eth_\circ^2 V_\circ(-n_i),\quad \sigma_i=- V_\circ^{-1}|\eth_\circ^2V_\circ|^2(-n_i).
\end{equation}
Indeed, the second term on the right hand side of~\eqref{eq:26} may be written as 
\begin{multline}
  \label{eq:27}
  \eth_\circ^2\widebar\eth_\circ^2 V_\circ(-n_i)=\eth_\circ[\eth_\circ \widebar\eth_\circ^2 V_\circ(-n_i)]-\eth_\circ \widebar\eth_\circ^2 V_\circ\eth_\circ(-n_i)\\=\eth_\circ[\eth_\circ \widebar\eth_\circ^2 V_\circ(-n_i)]-\eth_\circ[\widebar\eth_\circ^2 V_\circ\eth_\circ(-n_i)],
\end{multline}
which holds because $\eth_\circ^2(-n_i)=0$: $-n_i$ are combinations of harmonics with
$j=1$ which are in the kernel of $\eth_\circ^2$. 

The functional defined as
\begin{equation}
  \label{eq:102}
  \Phi_G[V_\circ]= \int  \frac{d^2\Omega}{4\pi}\ \phi_G, \quad \phi_G=  V^{-2}_\circ,
\end{equation}
corresponds to the volume associated to the metric determined by $V_\circ$
divided by $4\pi$. Its density satisfies an exact balance law,
\begin{equation}
  \label{eq:103}
  \partial_u \phi_G +\eth_\circ \mathcal J^G_s+\widebar \eth_\circ \widebar{\mathcal J}^G_s=0,
\end{equation}
with a current given by
\begin{equation}
  \label{eq:106}
  \mathcal J^G_s=\frac 13 \big(\eth_\circ V_\circ\widebar\eth_\circ^2V_\circ -V_\circ \eth_\circ \widebar\eth_\circ^2V_\circ  \big). 
\end{equation}
As a consequence, the RT evolution is volume preserving, 
\begin{equation}
  \label{eq:36}
  \partial_u\Phi_G=0,
\end{equation}
and $\Phi_G$ is constant. Furthermore, this functional is a potential for the
energy density in the sense that
\begin{equation}
  \label{eq:7}
  \delta (\frac{{\Phi_G}}{2})=-\int \frac{d^2\Omega}{4\pi}\ \varepsilon\ \delta V_\circ. 
\end{equation}

\subsection{Lyapunov function and total energy}
\label{sec:lyapunov-function}

In terms of the field $U^\infty_\circ(u,x)$ of spin and boost weights $0,1$ such that
\begin{equation}
  \label{eq:19}
  U^\infty_{\circ}=\int^\infty_u dv\ V_{\circ},
\end{equation}
the RT equation may be written as
\begin{equation}
  \label{eq:21}
  \partial_u \Psi^\infty_\circ=-V_\circ^{-1}|\eth_\circ^2 V_\circ|^2,\quad 
\end{equation}
where the generalized Bondi supermomentum
\begin{equation}
  \label{eq:22}
  \Psi^\infty_\circ=V_\circ^{-3}+\eth_\circ^2\widebar \eth_\circ^2U^\infty_{\circ},
\end{equation}
provides a local Lyapunov function for the RT equation. We will not discuss in
detail here convergence questions of these integrals. Note however that in the
explicit expression for the generalized Bondi supermomentum, only the higher
harmonics with $j\geq 2$ are relevant. 

Defining the non-linear memory density on the retarded time-interval $[u_1,u_2]$ as
\begin{equation}
  \label{eq:113}
  \mu^{u_2}_{u_1}=\int_{u_1}^{u_2} dv\, V_\circ^{-1}|\eth^2_\circ V_\circ|^2, 
\end{equation}
the future or accumulated non-linear memory density of the system is
\begin{equation}
  \label{eq:2}
  {\mu}^\infty_u=\int^\infty_udv\, V_\circ^{-1}|\eth^2_\circ V_\circ|^2. 
\end{equation}
In these terms, the RT equation is equivalent to the point-wise energy balance
equation,
\begin{equation}
  \label{eq:12}
  \partial_u\Psi^\infty_\circ=\partial_u \mu^\infty_u.
\end{equation}

As discussed in~\cite{Barnich:2026wpw}, if initial conditions for the RT
equation are set at $u=0$, another natural choice for an auxiliary variable is
\begin{equation}
  \label{eq:111}
  U_\circ^0=\int^u_0dv\, V_\circ, \quad 
\end{equation}
in terms of which generalized Bondi supermomentum is
\begin{equation}
  \label{eq:112}
  \Psi^0_\circ=V_\circ^{-3}-\eth_\circ^2\widebar \eth_\circ^2U^0_{\circ},\quad  \partial_u \Psi^0_\circ=-V_\circ^{-1}|\eth_\circ^2 V_\circ|^2, 
\end{equation}
while the RT equation may be written as a balance equation in terms of past accumulated non-linear memory density $\mu_0^u$ as
\begin{equation}
  \label{eq:114}
  \partial_u \Psi^0_\circ=-\partial_u \mu_0^u, 
\end{equation}
where the total non-linear memory density is
\begin{equation}
  \label{eq:127}
  \mu^\infty_0=\mu_0^u+{\mu}^\infty_u. 
\end{equation}

If one defines
\begin{equation}
  \label{eq:141}
  U^{\prime \infty}_\circ(u^\prime,x)=\int^\infty_{u^\prime} dv^\prime\, V^\prime_\circ (v^\prime,x), 
\end{equation}
this field transforms under scalings as 
\begin{equation}
  \label{eq:142}
  U^{\prime \infty}_\circ(u^\prime,x)=\lambda^3 U^{\infty}_\circ(\lambda^{-4}u^\prime,x). 
\end{equation}
Similarly, $U^0_\circ,\mu^\infty_u,\mu^u_0$ also transform with a multiplicative factor $\lambda^3$ under the scaling transformation. 

\subsection{Equilibrium states and thermodynamics}
\label{sec:equilibrium-states}

The equations
\begin{equation}
  \label{eq:120}
  \eth_\circ^2 V_\circ=0=\widebar\eth_\circ^2 V_\circ
\end{equation}
imply that $V_\circ$ is given by the first four spherical harmonics. In turn
this implies $\eth_\circ^2\widebar\eth_\circ^2 V_\circ=0$ and then also that the
harmonics are time independent on account of the RT equation,
\begin{equation}
  \label{eq:121}
  V^E_\circ=w_v=c^{jm}\
  {}_{0,1} Y_{jm}\rvert_{j\leq 1}.
\end{equation}
Alternative useful parametrizations of the equilibrium solutions are in terms of
4 real parameters $R,u_i$, or in terms of $R$ and the complex parameters $a,b,c,d$ with
$ad-bc=1$,
\begin{equation}
  \label{eq:123}
  w_v\equiv w_{R,g}=R^{-1}(u_0-u_i n^i)=R^{-1}\frac{|a\zeta+b|^2+|c\zeta+d|^2}{|\zeta|^2+1}. 
\end{equation}
The future directed unit vector $u^au_a=1$,
        \begin{equation}
          \label{eq:56}
          u_0(u_i)=\sqrt{1+|\vec u|^2}
        \end{equation}
        maybe also be parametrized as
  \begin{equation}
    \label{eq:243}
    \begin{split}
      u^0&=\frac{|a|^{2} + |c|^2+ |b|^{2}+|d|^{2}}{2}=\gamma=\cosh\omega,\\
      u^2&=\frac{\imath(a\widebar b+c\widebar d-b\widebar a-d\widebar c)}{2}=\gamma\beta^2=\sinh\omega\hat\omega^2,\\
      u^3&=\frac{|a|^{2}+|c|^{2}-|b|^{2}-|d|^{2}}{2}=\gamma\beta^3=
           \sinh\omega\hat\omega^3,\\
      u^1&=\frac{a\widebar b+c\widebar d
           +b\widebar a+d\widebar c}{2}=\gamma\beta^1=\sinh\omega\hat\omega^1,
    \end{split}
  \end{equation}
  while 
  \begin{equation}
    \label{eq:10} \begin{split} ({}^{R}Y_{j,-1},{}^{R}Y_{j0},{}^{R}Y_{j1})&
                                                                  =\sqrt{\frac{3}{4\pi}}\frac{1}{|\zeta|^2+1}(\imath(\widebar\zeta-\zeta),|\zeta|^2-1,\zeta+\widebar\zeta)\\ &=\sqrt{\frac{3}{4\pi}}(\sin\theta\sin\phi,\cos\theta,\sin\theta\cos\phi)         \\& =\sqrt{\frac{3}{4\pi}}(n^2,n^3,n^1).
    \end{split}
  \end{equation}
It follows that the energy is the monopole moment of the energy density in the equilibrium state. 
\begin{equation}
  \label{eq:124}
  E=P_0=\int  \frac{d^2\Omega}{4\pi}\ w_v^{-3}= R^3 u_0(u_i).
\end{equation}
We will also need the dipole moment (which from the gravitational viewpoint is
interpreted as the components of linear momentum of a boosted Schwarzschild
black hole),
\begin{equation}
  \label{eq:125}
  P_i=\int  \frac{d^2\Omega}{4\pi} w_v^{-3}(-n_i)= R^3 u_i, 
\end{equation}
where the mass of the black hole is given by 
\begin{equation}
  \label{eq:126}
  M= R^3. 
\end{equation}

Gravitational considerations~\cite{Barnich:2026dka} imply that the entropy,
defined as the area of the horizon, is
\begin{equation}
  \label{eq:8}
  S=4\pi M^2. 
\end{equation}
In these considerations, the volume of the equilibrium metrics, 
\begin{equation}
  \label{eq:4}
4\pi \Phi_G[w_v]= \int d^2\Omega\ w^{-2}_v=4\pi R^2,
\end{equation}
turns out to be important.

From the variation of the fundamental relation $S(P_a)$ in the entropy
representation,
\begin{equation}
  \label{eq:128}
  \delta S=\beta^a\delta P_a, 
\end{equation}
it then follows that
\begin{equation}
  \label{eq:129}
  \beta^a =8\pi M u_a=8\pi P^a \iff P_a=\frac{1}{8\pi}\beta_a,
\end{equation}
while the fundamental relation and the Massieu function are given by
\begin{equation}
  \label{eq:130}
  S(P_a)=4\pi P_a P^a,\quad \Phi(\beta^a)=[S-\beta^aP_a]|_{P=P(\beta)}=-\frac{1}{16\pi}\beta^a\beta_a. 
\end{equation}

{\bf Remarks:}

(i) The fundamental relation scales as a homogeneous function of degree
$2$ rather than the more usual scaling of degree $1$, $S(\Lambda P_a)=\Lambda^2S(P_a)$, with
$\Lambda=\lambda^3$ when comparing to the scaling in~\eqref{eq:138}. As a consequence, the equations of
state are homogeneous functions of degree $1$ rather than degree $0$,
\begin{equation}
  \label{eq:47}
  \beta^a(\Lambda P_b)=\Lambda \beta^a(P_b). 
\end{equation}
When choosing $\Lambda=1/M=1/R^3$, and defining $p_a=P_a/M$ with $p^ap_a=1$,
$b^a=\beta^a/M$, this may be used to remove $M$, respectively $R$, as an
independent variable in the usual way. The independent parameters for the reduced
problem are $u_i$, with associated conjugate variables $p_i,b^i$.
Writing $S(P_a)=4\pi M^2s(p_i)$, the gravitational result implies that the
reduced entropy is trivial, $s(p_i)=1$.

(ii) Note that if one would consider as non-equilibrium Massieu functional simply
\begin{equation}
  \label{eq:32}
  \Phi_M[V_\circ]=\frac 12 \Phi_G, 
\end{equation}
it follows from equation~\eqref{eq:7} that
\begin{equation}
  \label{eq:55}
  \delta \Phi_M=-\int \frac{d^2\Omega}{4\pi}\ \varepsilon \delta V_\circ, \quad \varepsilon=V_\circ^{-3}
\end{equation}
This means that one takes as energy density the Bondi mass aspect while the
field $V_\circ$ plays the role of local inverse temperature $\beta_{\mathrm{loc}}$. This
functional is conserved on account of~\eqref{eq:103}. At equilibrium,
$\Phi_M[w_v]=\frac 12 R^2$, which does not agree with the equilibrium
thermodynamics of boosted Schwarzschild black holes. This set-up does not seem
to be entirely natural or satisfactory from the current perspective. There would
be Bondi mass-loss but no entropy production.

\section{Local Gibbs relation and non-equilibrium entropy functional}\label{sec:non-equil-entr}

\subsection{Interpretation from the viewpoint of the gravitational system}\label{sec:interpr-from-viewp}

Comparing the balance equation~\eqref{eq:12} to a local version of~\eqref{eq:59} suggests to define,
by analogy,
\begin{equation}
  \label{eq:67}
  \begin{split}
    & \mathrm{internal\ energy\ density}\ \varepsilon(=u)=V_\circ^{-3},\\
    & \mathrm{kinetic\ energy\ density}\ k^\infty =\eth^2_\circ\widebar\eth_\circ^2 U^\infty_\circ,\
      \partial_u k^\infty =\eth_\circ\mathcal J_e+\widebar\eth_\circ\widebar{\mathcal J_e}, \\
    & \mathrm{rate\ of\ heat\ density\ loss}\ \dot q^\infty= -V_\circ^{-1} |\eth_\circ^2 V_\circ|^2, \end{split}                                                                                 
\end{equation}
so that equation~\eqref{eq:12} now reads
\begin{equation}
  \label{eq:148}
  \partial_u \varepsilon + \partial_u k^\infty=\dot q^\infty. 
\end{equation}
In particular, this identifies the opposite of non-linear memory density,
$-V_\circ^{-1} |\eth_\circ^2 V_\circ|^2$ for RT spacetimes, with the rate of heat density loss.
Note that this identification may also be done more generally in locally
asymptotically flat spacetimes at future null infinity in the 
Bondi-Metzner-Sachs or Newman-Penrose framework.

For the monopole contribution, the internal energy is the Bondi mass aspect
\begin{equation}
  \label{eq:72}
  P_0(=U)=\int \frac{d^2\Omega}{4\pi}\ V_\circ^{-3},
\end{equation}
the kinetic energy does not contribute, and the future accumulated heat loss
corresponds to the integrated, future accumulated non-linear memory,
\begin{equation}
  \label{eq:73}
  Q^\infty=\int \frac{d^2\Omega}{4\pi}\ \mu_u^\infty=\int^\infty_u dv \int \frac{d^2\Omega}{4\pi}\ V_\circ^{-1}|\eth_\circ^2 V_\circ|^2,
\end{equation}
so that the integrated first law takes the form
\begin{equation}
  \label{eq:74}
  \frac{d}{du} P_0=\dot Q^\infty,
\end{equation}
with
$\dot Q^\infty=-\int \frac{d^2\Omega}{4\pi}\ V_\circ^{-1} |\eth_\circ^2 V_\circ|^2$ the
rate of total heat loss. Similarly, for the dipole contribution, kinetic energy
does not contribute either, and the balance equation is
\begin{equation}
  \label{eq:75}
 \frac{d}{du} P_i=\int \frac{d^2\Omega}{4\pi}\ V_\circ^{-1}|\eth_\circ^2 V_\circ|^2(n_i),  
\end{equation}
while for the higher moments $j\geq 2$ of the generalized Bondi supermomentum
\begin{equation}
  \label{eq:52}
  P^\infty_{jm}=\int d^2\Omega\, \Psi^\infty_\circ\, {}_{0,-3} \widebar Y_{jm},
\end{equation}
kinetic energy can no longer be neglected and
\begin{equation}
  \label{eq:76}
  \frac{d}{du} P_{jm}^\infty=-\int d^2\Omega\, V_\circ^{-1}|\eth_\circ^2 V_\circ|^2\ {}_{0,-3}\widebar Y_{jm}. 
\end{equation}

If one now defines the non-equilibrium entropy density as
\begin{equation}
  \label{eq:84}
  s^{\infty}=\int^\infty_u dv\ |\eth_\circ^2 V_\circ|^2,
\end{equation}
and the local version of inverse temperature as
\begin{equation}
  \label{eq:85}
  \beta_{\mathrm{loc}}=V_{\circ}, 
\end{equation}
so that
\begin{equation}
  \label{eq:86}
  S^{\infty}[V_\circ]=\int^\infty_u dv \int \frac{d^2\Omega}{4\pi}\ |\eth_\circ^2 V_\circ|^2, \quad
  \frac{d S^{\infty}}{du}=
  \int \frac{d^2\Omega}{4\pi}\ \beta_{\mathrm{loc}}\ \dot q\leq 0. 
\end{equation}
As a consequence, the RT equation in the form~\eqref{eq:104} is equivalent to a local
version of the Gibbs relation~\eqref{eq:64}, without a work term,
\begin{equation}
  \label{eq:96}
  \frac{d s^{\infty}}{du}=\beta_{\mathrm{loc}}\big(\frac{d\varepsilon}{du}+\frac{dk^\infty}{du}\big). 
\end{equation}
Note that we have been using the equality sign in~\eqref{eq:61} appropriate for
equilibrium processes. The entropy is decreasing in agreement with our
interpretation that the gravitational system is cooling down and transfering
heat to the environment.

\subsection{Interpretation from the viewpoint of the environment}\label{sec:interpr-from-viewp-1}

An alternative interpretation from the viewpoint of the environment, or perhaps
more concretely, from the viewpoint of an asymptotic time-like observer with its
interferometer measuring gravitational radiation, proceeds as follows,
\begin{equation}
  \label{eq:134}
  \begin{split}
    & \mathrm{kinetic\ energy\ density}\ k^0 =\eth^2_\circ\widebar\eth_\circ^2 U^0_\circ, \
      \partial_u k^0 =-\eth_\circ\mathcal J_e-\widebar\eth_\circ\widebar{\mathcal J_e},\\
    & \mathrm{heat\ density\ gain}\ \dot q^0= V_\circ^{-1} |\eth_\circ^2 V_\circ|^2, \\
    & \mathrm{internal\ energy\ density\ of\ the\ external\ system}\ \varepsilon=V_\circ^{-3},
  \end{split}
\end{equation}
and the local energy balance becomes 
\begin{equation}
  \label{eq:135}
  \dot q^0=\partial_u k^0-\partial_u \varepsilon,
\end{equation}
that is to say the rate of heat gained by the environment equals the kinetic
energy of radiation plus the loss of internal energy density by the system. In
this case, the total past accumulated heat gained by the environment corresponds
to the past accumulated non-linear memory,
\begin{equation}
  \label{eq:133}
  Q^0=\int\frac{d^2\Omega}{4\pi}\, \mu^u_0,  
\end{equation}
while the local entropy density becomes
\begin{equation}
  \label{eq:136}
  s^{0}=\int^u_0dv\, |\eth_\circ^2 V_\circ|^2,
\end{equation}
with the local version of temperature still being $\beta_{\mathrm{loc}}=V_\circ$. The non
equilibrium entropy functional becomes
\begin{equation}
  \label{eq:137}
  \boxed{S^{0}[V_\circ]=\int^u_0 dv \int \frac{d^2\Omega}{4\pi}\ |\eth_\circ^2 V_\circ|^2}, \quad
  \frac{d S^{0}}{du}=
  \int \frac{d^2\Omega}{4\pi}\ \beta_{\mathrm{loc}}\ \dot q\geq 0, 
\end{equation}
while the local version of the Gibbs relation now becomes
\begin{equation}
  \label{eq:139}
   \boxed{\frac{d s^{0}}{du}=\beta_{\mathrm{loc}}\big(\frac{dk^0}{du}-\frac{d\varepsilon}{du}\big)}.
 \end{equation}

\section{Combining with the equilibrium sector}
\label{sec:comb-with-equil}

\subsection{Equilibrium contribution}\label{sec:equil-contr}

One may define an additional equilibrium contribution in terms of the state
variable $V_\circ$ through the constant functional $\Phi_G$,
\begin{equation}
  \label{eq:63}
  S^{\mathrm{eq}}[V_\circ]=4\pi \Phi_G^3.
\end{equation}
In particular, the non-equilibrium contribution vanishes at equilibrium,
$S^{0}[w_v]=0$ while the equilibrium contribution correctly reproduces the
equilibrium result, $S^{eq}[w_v]=4\pi M^2$. A remaining question is then how those
two should be combined.

One possibility would be to just consider their sum and define the total entropy
through
\begin{equation}
  \label{eq:109}
  S^{\mathrm{tot}}[V_\circ]=S^{\mathrm{eq}}+S^{0}.
\end{equation}
This proposal fulfills the primary aim to define an entropy functional $S^{tot}[V_\circ]$
for the RT equation in terms of the state variable $V_\circ$ that reduces to the
equilibrium entropy. 

\subsection{Reduced scale invariant description}\label{sec:reduc-scale-invar}

There might however be other considerations to take into account. The first term
in the sum above scales as $\lambda^{6}$, while the second term scales as
$\lambda^2$. If one requires the entropy to scale as $\lambda^6$, as it
does in the equilibrium theory,
\begin{equation}
  \label{eq:39}
  S[\lambda^{-1}V_\circ(\lambda^{-4}u,x)]=\lambda^6S[V_\circ(u,x)], 
\end{equation}
one may use the functional $\Phi^G$ that is a constant and scales as
$\lambda^2$ to define
\begin{equation}
  \label{eq:149}
  S=4\pi \Phi_G^3+4\pi \Phi_G^2S^{0}. 
\end{equation}
Alternatively, the same result may be arrived at by setting 
\begin{equation}
  \label{eq:108}
  \lambda^{-1}=\Phi_G^{\frac 12}. 
\end{equation}
and consider the reduced, scale invariant variable
\begin{equation}
  \label{eq:46}
  v_\circ=(\Phi^G)^{\frac 12}V_\circ,
\end{equation}
which has a trivial monopole contribution for the associated volume, 
\begin{equation}
  \label{eq:51}
  \int \frac{d^2\Omega}{4\pi}\ v_\circ^{-2}=1.
\end{equation}
As discussed in~\eqref{eq:138}, defining $u^\prime=(\Phi^G)^{-2}u$, the variable
$v_\circ(u^\prime,x)=\Phi_G^{\frac 12}V_\circ(\Phi_G^2u^\prime,x)$ satisfies the evolution equation~\eqref{eq:77}
with $V_\circ(u,x)$ replaced by $v_\circ(u^\prime,x)$ and $u$ replaced by $u^\prime$. The standard
scaling argument then implies that entropy may be written in terms of a reduced
entropy as
\begin{equation}
  \label{eq:48}
  S[V_\circ]=4\pi \Phi_G^3S^{0}_R[v_\circ],\quad S^{0}_R[v_\circ]=1+\int_0^{\Phi_G^{-2}u}dv^\prime \int
  \frac{d^2\Omega}{4\pi}\ |\eth_\circ^2 v_\circ(v^\prime, x)|^2. 
\end{equation}
At equilibrium $v_\circ$ is given by
$w_{1,g}$, and $S^{0}_R[w_{1,g}]=1$.

Two other avenues to combine the two sectors may be pursued. The first is to
follow~\cite{rendall1988existence}: one fixes $\Phi_G=R^2=M^{2/3}$ and considers the flow for $v_\circ^{-2}$
with antipodally symmetric initial data. In this case, there are no low
harmonics and exponential decrease towards a single Schwarzschild black hole
with mass $M$ can be established. The equilibrium sector together with its
thermodynamics becomes trivial.

The case with generic initial data is more interesting. In order to study
convergence, an adapted version of the de Turck trick has been used
in~\cite{Singleton_1990,Singleton2020,Chrusciel:1992cj,Chrusciel:1992rv,Chrusciel:1992tj}. This allows one to better separate the
dynamics of the low harmonics from that of the convergent higher harmonics. It
explicitly involves dynamical equations for the homogeneous space
$\mathbb R^+_\star \times SL(2,\mathbb C)/SU(2)$, that captures the equilibrium
solutions. This procedure is also relevant in the context of perturbation
theory~\cite{Barnich:2026wpw}. It would be interesting to study in more detail what one may
learn from this approach about the interplay between the equilibrium and 
non-equilibrium sectors.

\section{Phase space considerations}\label{sec:phase-space-its}

\subsection{Lorentz invariant decomposition of phase space}\label{sec:lorentz-invar-decomp}

For the RT equation, one may introduce conjugate momenta as follows. Let
$\mathcal P(u,x)$ be a new independent real field of spin and boost weights
$(0,-3)$. One may define the canonical equal-time Poisson brackets
\begin{equation}
  \label{eq:116}
  \{V_\circ(u,x),\mathcal P(u,x^\prime)\}=\delta(x,x^\prime), \quad \{V_\circ(u,x),V_\circ(u,x^\prime)\}=0=\{\mathcal P(u,x),\mathcal P(u,x^\prime)\},
\end{equation}
where
\begin{equation}
  \label{eq:14}
 d^2\Omega=\frac{d\zeta d\widebar\zeta}{\imath P^2_\circ},\quad \delta(x,x^\prime)\equiv \delta(\zeta,\widebar\zeta;\eta,\widebar\eta)=\imath  P^2_\circ\delta(\zeta-\eta)\delta(\widebar\zeta-\widebar\eta).
\end{equation}
Under the scaling transformation, we define
\begin{equation}
  \label{eq:117}
  \mathcal P^\prime(u^\prime,x)= \lambda^3 \mathcal P(\lambda^{-4}u^\prime,x).
\end{equation}

The fields may also be expanded into (spin and boost weighted) spherical harmonics as follows,
\begin{equation}
  \label{eq:63a}
  V_\circ=q^{jm}(u)\ {}_{0,1}Y_{jm},\quad \mathcal P=p_{jm}(u)\ {}_{0,-3} Y_{jm}
\end{equation}
with $\widebar q^{jm}=(-)^m q^{j,-m}$, $\widebar p_{jm}=(-)^mp_{j,-m}$, and the symplectic potential becomes 
\begin{equation}
  \label{eq:119}
  \int d^2\Omega\ \dot V_\circ  \mathcal P=\dot q^{jm}\widebar p_{jm}. 
\end{equation}

One may perform an orthogonal decomposition of phase space in terms of low and
high harmonics as follows:
\begin{equation}
  \label{eq:16}
  V_\circ=V_{\circ j\leq 1}+V_{\circ j>1},\quad \mathcal P={\mathcal P}_{j\leq 1}+{\mathcal P}_{j> 1}, 
\end{equation}
The space $V_{\circ j\leq 1}$ is the four dimensional Lorentz invariant subspace of the
spherical harmonics of spin and boost weights $(0,1)$ selected by equations~\eqref{eq:120}.
It is orthogonal to the infinite dimensional Lorentz invariant subspace
${\mathcal P}_{j> 1}$ of the spherical harmonics of spin and boost weights
$0,-3$ (see~e.g.~\cite{Penrose:1984}, chapter 4 and recent considerations
in~\cite{Bekaert:2024uuy,Bekaert:2025kjb}). Elements of ${\mathcal P}_{j> 1}$
may be written as,
\begin{equation}
  \label{eq:17}
  \mathcal P_{j>1}=\eth^2_\circ\widebar\eth^2_\circ U, 
\end{equation}
where $U$ is of spin and boost weights $0,1$, and scales as
\begin{equation}
  \label{eq:81}
  U^\prime(u^\prime,x)=\lambda^{3}U(\lambda^{-4}u^\prime,x).
\end{equation}
The orthogonal decomposition of the symplectic potential is 
\begin{equation}
  \label{eq:23}
  \int d^2\Omega\ \dot V_\circ  \mathcal P=\int d^2\Omega\ [\dot V_{\circ j\leq 1}  \mathcal P_{j\leq 1}+\dot V_{\circ j>1}  \mathcal P_{j> 1}].
\end{equation}
Note that the quotient representation $\mathcal P/\mathcal P_{j>1}$ is
isomorphic to the representation $V_{\circ j\leq 1}$.

The degrees of freedom associated to the low harmonics may be described in terms
of real harmonics, $x^a(u)$, $p_a(u)$. If
$(\sin\theta\sin\phi,\cos\theta,\sin\theta\cos\phi) =(n^2,n^3,n^1)$ and
\begin{equation}
  \label{eq:24}
  V_{\circ j\leq 1}=x^0-x^i n_i,\quad \mathcal P_{j\leq 1}=\frac {1}{4\pi} (p_0+3 p_i n^i),
\end{equation}
it follows that
\begin{equation}
  \label{eq:28}
  \int d^2\Omega\ \dot V_{\circ j\leq 1}  \mathcal P_{j\leq 1}=\dot x^a p_a. 
\end{equation}
We will refer to these degrees of freedom as ``particle/black hole'' degrees of
freedom. They are the only relevant ones at equilibrium where they become
time-independent.

The ``radiative'' degrees of freedom are those associated to the higher harmonics
with the symplectic potential given by
\begin{equation}
  \label{eq:29}
  \int d^2\Omega\ \dot V_{\circ j> 1}  \eth^2_\circ\widebar\eth^2_\circ U. 
\end{equation}
In the presence of radiation, the dynamics of the particle degrees of freedom
becomes non-trivial. This point of view is related to the interpretation
developed in~\cite{newman1969equations,Newman1969}. 

\subsection{Constitutive relations}
\label{sec:const-relat}

Consider a formulation with the fields $V_\circ(u,x),U^\infty_\circ(u,x)$ respectively
$U^0_\circ(u,x)$ and their conjugate momenta $P_G,P_R$, with non-vanishing Poisson
brackets
\begin{equation}
  \label{eq:3}
  \{V_\circ(u,x),P_G(u,x^\prime)\}=\delta(x,x^\prime)=  \{U^\infty_\circ(u,x),P_R(u,x^\prime)\}, 
\end{equation}
or the analog relations with $U^\infty_\circ$ replaced by $U^0_\circ$ and 
the functionals
\begin{equation}
  \label{eq:18}
  \Phi_G= \int \frac{d^2\Omega}{4\pi}\, V_\circ^{-2},\quad  \Phi_R=-\int \frac{d^2\Omega}{4\pi}\, |\eth^2_\circ U^\infty_{\circ}|^2,\quad
  \Phi= \frac 12 (\Phi_G+\Phi_R),
  \end{equation}
  respectively
  \begin{equation}
    \label{eq:68}
    \Phi_R=\int \frac{{d^2\Omega}}{4\pi}\, |\eth^2_\circ U^0_{\circ}|^2,
  \end{equation}
and impose the ``constitutive'' relations
\begin{equation}
  \label{eq:20}
  P_G(V_\circ)=-\frac{\delta (4\pi \Phi)}{\delta V_\circ}=V_\circ^{-3},\, P_{R}(U^\infty_{\circ})
  =-\frac{\delta (4\pi \Phi)}{\delta U^\infty_{\circ}}=\eth_\circ^2\widebar\eth_\circ^2U^\infty_{\circ},
\end{equation}
respectively
\begin{equation}
  \label{eq:69}
  P_{R }(U^0_{\circ})=-\frac{\delta (4\pi\Phi)}{\delta U^0_{\circ}}=-\eth_\circ^2\widebar\eth_\circ^2U^0_{\circ}, 
\end{equation}
see Appendix~\ref{sec:dirac-brack-equil} for an interpretation in terms of first class constraints and a
Legendre transform. Note that the relations are invertible provided one
restricts to the higher modes, $P_{R j >1}$, $U^\infty_{\circ j >1}$, $U^0_{\circ j>1}$ and
that they are compatible with the required scaling.

    It then follows that the functional $\Phi$ is a generating functional for
    generalized Bondi super-momentum in the sense that
\begin{equation}
  \label{eq:83}
  \Psi^\infty_\circ=P_G(V_\circ)+P_{R j>1}(U^\infty_{\circ j>1}),
\end{equation}
respectively
\begin{equation}
  \label{eq:79}
  \Psi^0_\circ=P_G(V_\circ)+P_{R j>1}(U^0_{\circ j>1}).
\end{equation}

\subsection{Non-dissipative RT equation as a constrained Hamiltonian system}
\label{sec:non-dissipative-rt}

The ``non-radiative'' RT
equation may be defined as the evolution equation~\eqref{eq:77} without the second
term on the RHS,
\begin{equation}
  \label{eq:78}
  3\partial_u V_\circ=-V_\circ^4\eth_\circ^2\widebar\eth_\circ^2 V_\circ\iff \partial_u V_\circ^{-3}-\eth^2_\circ\widebar\eth_\circ^2 V_\circ=0. 
\end{equation}
Bondi supermomentum $\Psi^\infty_\circ,\Psi^0_\circ$ as defined in~\eqref{eq:22}
or~\eqref{eq:112}, provides an infinite number of conservation laws for this
evolution equation. 

The first class Hamiltonian
\begin{equation}
  \label{eq:70}    H=\int d^2\Omega \Big[-(P_R-\eth_\circ^2\widebar\eth_\circ^2 U^\infty_\circ)V_\circ-\frac 13(P_G-V_\circ^{-3})V_\circ^4\eth^2_\circ\widebar\eth_\circ^2 V_\circ\Big],
\end{equation}
respectively
\begin{equation}
  \label{eq:80}
  H=\int d^2\Omega \Big[(P_R+\eth_\circ^2\widebar\eth_\circ^2 U^0_\circ)V_\circ-\frac 13(P_G-V_\circ^{-3})V_\circ^4\eth^2_\circ\widebar\eth_\circ^2 V_\circ\Big],
\end{equation}
which is a combination of the first class constraints, generates the correct evolution equations, 
\begin{equation}
  \label{eq:71}
  \begin{split}
    \dot U^\infty_\circ&=\frac{\delta H}{\delta P_R}=-V_\circ,\quad (\dot U^0_\circ
                                                              =\frac{\delta H}{\delta P_R}=V_\circ)\\
     \dot P_R&=-\frac{\delta H}{\delta U^\infty_\circ}=-\eth^2_\circ\widebar\eth^2_\circ V_\circ,\quad (\dot P_R=-\frac{\delta H}{\delta U^0_\circ}=-\eth^2_\circ\widebar\eth^2_\circ V_\circ)\\
  \dot V_\circ&=\frac{\delta H}{\delta P_G}=-\frac 13 V_\circ^4\eth^2_\circ\widebar\eth_\circ^2 V_\circ,\\
    \dot P_G&=-\frac{\delta H}{\delta V_\circ}=(P_R-\eth^2_\circ\widebar\eth^2_\circ U^\infty_\circ)+\frac 13\eth^2_\circ\widebar\eth^2_\circ[(P_G-V_\circ^{-3})V_\circ^4]\\ & \hspace{5cm}+\frac{4}{3}[(P_G-V_\circ^{-3})V_\circ^3]\eth^2_\circ\widebar\eth^2_\circ V_\circ+\eth^2_\circ\widebar\eth^2_\circ V_\circ,\\
    (\dot P_G&=-\frac{\delta H}{\delta V_\circ}=-(P_R+\eth^2_\circ\widebar\eth^2_\circ U^0_\circ)+\frac 13\eth^2_\circ\widebar\eth^2_\circ[(P_G-V_\circ^{-3})V_\circ^4]\\ & \hspace{5cm}+\frac{4}{3}[(P_G-V_\circ^{-3})V_\circ^3]\eth^2_\circ\widebar\eth^2_\circ V_\circ+\eth^2_\circ\widebar\eth^2_\circ V_\circ),
  \end{split}
\end{equation}
in the sense that the evolution equations for the momenta are consistent with
the constraints. 

\subsection{Adding dissipation}\label{sec:adding-dissipation}

For the full RT equation~\eqref{eq:77} with the radiative term, instead of
conserved generalized Bondi supermomentum, one finds strictly positive
supermomentum loss as discussed previously, while the last two Hamiltonian
equations are modified as
\begin{equation}
  \label{eq:88}
  \begin{split}
    \dot V_\circ&=\frac{\delta H}{\delta P_G}+\frac 13 V_\circ^3|\eth^2_\circ V_\circ|^2, \\
    \dot P_G & =-\frac{\delta H}{\delta V_\circ}-V_\circ^{-1}|\eth^2_\circ V_\circ|^2.
  \end{split}
\end{equation}
It remains to be seen whether such a formulation is useful in the context of
non-equilibrium thermodynamics or in an attempt at a microscopic formulation.

\section{Discussion}\label{sec:discussion}

The starting point for our approach here has been a first law for an
asymptotically time-like observer in an asymptotically flat spacetime at future
null infinity, whose energy density is the generalized Bondi mass aspect
associated to RT solutions. Proceeding by analogy, we then have identified heat
density, a local version of temperature and a non-decreasing entropy density
that can be used to recast the RT equation as a local Gibbs relation.

The hope is that these results might shed some light on earlier and recent
proposals for non-equilibrium thermodynamics of gravitational systems. From the
current perspective, the next step consists in understanding the considerations
here in terms of the various horizons associated to RT
spacetimes~\cite{Tod1989,Chrusciel:1992rv,Chrusciel:1992tj,Chow:1995va2,%
  Rezzolla:2010df,Jaramillo:2011re}.

Entropy currents and transport coefficients for RT spacetimes have been
discussed extensively in the context of both the AdS/CFT and the fluid gravity
correspondence~\cite{BernardideFreitas:2014eoi,Bakas:2014kfa,Skenderis:2017dnh,%
  Ciambelli:2017wou}. Since the effective two dimensional Calabi flow for
RT spacetimes is the same for all values of the cosmological constant, it should
be most instructive to understand in detail the relation and differences between
these paradigms (see e.g.~\cite{Arenas-Henriquez:2025rpt,Diaz:2026igh} for considerations along these
lines). A simple starting point could be to work out the equilibrium
thermodynamics of boosted AdS Schwarzschild black holes.

More generally, one might want to try to relate the entropy functional for RT
spacetimes constructed here to recent proposals for gravitational
non-equilibrium entropy proposals based on generic null hypersurfaces,
dynamical, quasi-local and isolated
horizons~\cite{RignonBret2023,Hollands:2024vbe,Ashtekar2026a,%
  Ashtekar2026}. Since the starting point, both for the equilibrium
and the non-equilibrium thermodynamics, has been the definition of energy as
seen from infinity in asymptotically flat spacetimes, a comparison requires to
complete our analysis with a detailed study of RT horizons as mentioned above.
Recent considerations on observer dependent constructions~\cite{Shajiee:2026cwg} and on the
generalized second law from the viewpoint of null infinity with quantum rather
than classical fields~\cite{RignonBret2026,RignonBret2026a} might also be very relevant.

\section*{Acknowledgements}\label{sec:acknowledgements}

The author is grateful to P.~Gaspard for important advice and to J.L.~Jaramillo
and M.~Lenzi for useful discussions. This work is supported by the F.R.S.-FNRS
Belgium through convention IISN 4.4516.26 and CDR grant J.0006.26.

\vfill
\pagebreak

\appendix

\section{First class constraints and Legendre transform}
\label{sec:dirac-brack-equil}

If $q^\alpha,p_\alpha$ are local coordinates on phase space, the canonical symplectic form
is $\omega=dp_\alpha\wedge dq^\alpha$ and the associated canonical Poisson brackets are
\begin{equation}
  \label{eq:30}
  \{q^\alpha,p_\beta\}=\delta^\alpha_\beta,\quad \{q^\alpha,q^\beta\}=0=\{p_\alpha,p_\beta\}.
\end{equation}
Consider a configuration space function $\Phi(q)$ with non-degenerate Hessian
matrix,
\begin{equation}
  \label{eq:31}
  H_{\alpha\beta}=\frac{\partial^2 \Phi}{\partial q^\alpha\partial q^\beta},\quad |H_{\alpha\beta}|\neq 0,
\end{equation}
and denote by $H^{\alpha\beta}$ its inverse, $H^{\alpha\beta}H_{\beta\gamma}=\delta^\alpha_\gamma$.

The constraints
\begin{equation}
  \label{eq:33}
  \pi_\alpha\equiv p_\alpha+\frac{\partial\Phi}{\partial q^\alpha}= 0,
\end{equation}
are first class since they satisfy
\begin{equation}
  \label{eq:34}
  \{\pi_\alpha,\pi_\beta\}=0.
\end{equation}
They generate the gauge transformations
\begin{equation}
  \label{eq:35}
  \delta_\epsilon q^\alpha=\{q^\alpha,\pi_\beta\epsilon^\beta\}=\epsilon^\alpha,\quad \delta_\epsilon p_\alpha=\{p_\alpha,\pi_\beta\epsilon^\beta\}=-H_{\alpha\beta}\epsilon^\beta, 
\end{equation}
and there are no dynamical degrees of freedom in the reduced phase space.

Assumption~\eqref{eq:31} guarantees equation~\eqref{eq:33} may be inverted to
yield $q^\alpha=q^\alpha(p)$. One may define
\begin{equation}
  \label{eq:13}
  \Phi^*(q^\alpha,p_\beta)=q^\alpha p_\alpha+\Phi(q), 
\end{equation}
so that the constraints are equivalent to 
\begin{equation}
  \label{eq:15}
  \frac{\partial \Phi^*}{\partial q^\alpha}=0, 
\end{equation}
while the Legendre transform $\hat\Phi(p)$ of $\Phi(q)$ with respect to $q^\alpha$ is
\begin{equation}
  \label{eq:37}
  \hat \Phi(p)=\big[\Phi^*(q^\alpha,p_\beta)\big]\big\rvert_{q=q(p)}.
\end{equation}

\vfill
\newpage

\addcontentsline{toc}{section}{References}

\printbibliography%

\end{document}